\documentclass[fleqn,twoside]{article}
\usepackage{espcrc2}

\usepackage{graphicx}
\usepackage[figuresright]{rotating}

\newcommand{\AmS}{{\protect\the\textfont2
  A\kern-.1667em\lower.5ex\hbox{M}\kern-.125emS}}

\title{Testing astroparticle physics with the Fermi Large Area Telescope}

\author{Aldo Morselli\address[a]{INFN Roma Tor Vergata } {\sl on behalf of the Fermi LAT collaboration} }%
       
\begin{document}

\begin{abstract}
Our understanding of the Universe today includes overwhelming observational evidence for the existence of an elusive form of matter that is generally referred to as dark. Although many theories have been developed to describe its nature, very little is actually known about its properties. 
Since its launch in 2008, the Large Area Telescope, onboard the  Fermi Gamma-ray Space Telescope, has detected by far the greatest number ever of gamma rays, in the 20MeV 300GeV energy range and electrons + positrons in the 7 GeV- 1 TeV range. This impressive statistics allows one to perform a very sensitive indirect experimental search for dark  matter. I will present the latest results on these searches.
\vspace{1pc}
\end{abstract}

\maketitle

\section{The Cosmic Ray Electron spectrum}
Recently the experimental information available on the Cosmic Ray Electron (CRE) spectrum has been dramatically expanded as the $Fermi$ LAT Collaboration~\cite{fermi} has reported a high precision measurement of the electron spectrum from 7 GeV to 1 TeV performed with its Large Area Telescope (LAT) \cite{Fermi_el}, \cite{Fermi_el2}. The spectrum  shows no  prominent spectral features and it is significantly harder than that inferred from several previous experiments.  These data together with the PAMELA data on the rise above 10 GeV of the positron fraction \cite{Pam_pos} are quite difficult to explain with just secondary production~\cite{SM134},\cite{jcap}, \cite{IDM08}.

\begin{figure}[p]
\vskip -0.4 cm
\includegraphics[width=8.2cm,height=7.8cm]{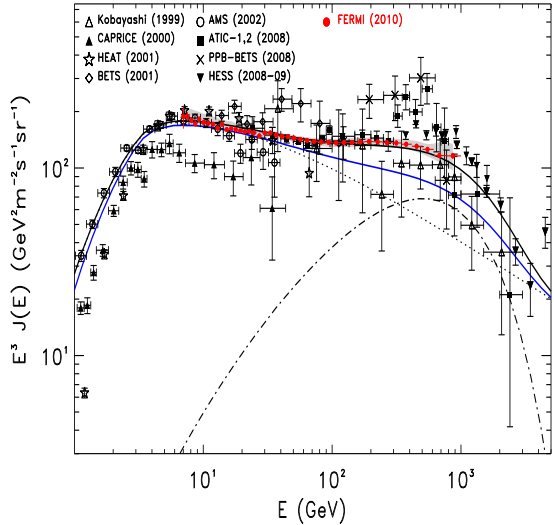}
\vskip -0.9 cm
\caption{\label{Fermi_el_fig} 
Cosmic ray electron + positron spectrum as measured by $Fermi$ Large Area Telescope  for one year of observations  (filled circles), along with other recent high energy results. 
The gray band represents systematic errors on the  $Fermi$ LAT data \cite{Fermi_el},  \cite{Fermi_el2}.
The solid line  is the computed  conventional GALPROP model  but with an  injection  index $\Gamma $= 1.6/2.7 below/above 4 GeV (dotted line). An additional component with an injection index $ \Gamma = 1.5$ and exponential cut-off is shown by the dashed line. 
Blue line shows $e^-$ spectrum only.
}
\end{figure}

\begin{figure}[!t]
 \centering
  \includegraphics[width=7.9cm,height=5.5cm ]{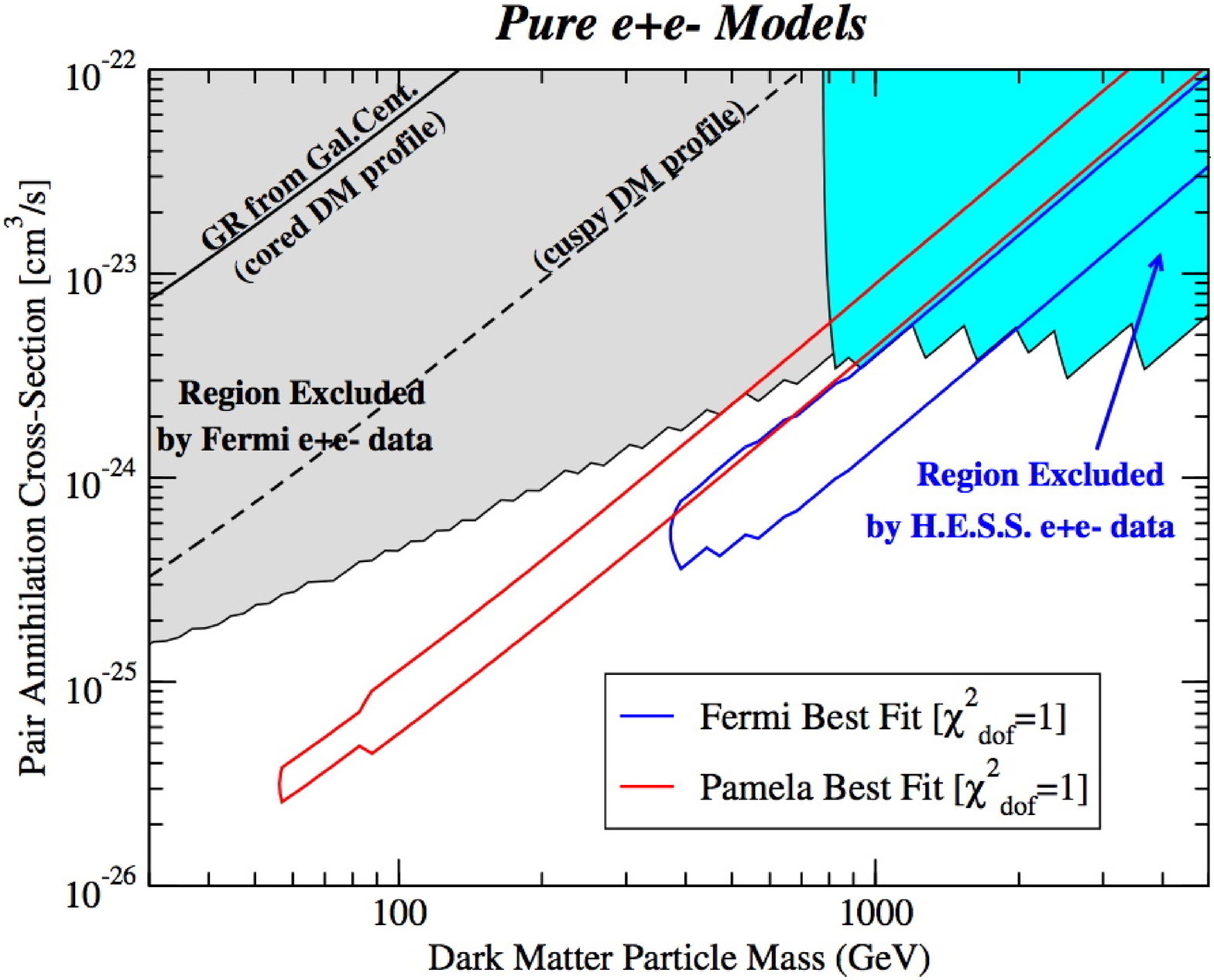}
\vskip -0.6 cm
 \caption{  The parameter space of dark matter particle mass versus pair-annihilation rate, for models where dark matter annihilates into monochromatic $e^\pm$ .    
 Models inside the regions shaded in gray and cyan over-produce $e^\pm$ from dark matter annihilation with respect to the $Fermi$ LAT and H.E.S.S. measurements, at 
 2-$\sigma$ level. The red and blue contours outline the regions where the $\chi^2$ per degree of freedom for fits to the PAMELA and $Fermi$ LAT data is less than 1. 
  }
\label{fig:mod1}
\vskip -0.6 cm
\end{figure}
The temptation to claim the discovery of dark matter  from detection of  electrons from annihilation of dark matter particles is strong
but  there are competing astrophysical sources, such as pulsars, that can give a strong 
flux of primary positrons and electrons
(see \cite{puls0}, \cite{puls}, \cite{coutu}, \cite{int_pap} and references therein).
At energies between 100 GeV and 1 TeV the electron flux reaching the Earth may be the sum of an almost homogeneous and isotropic component produced by  Galactic supernova remnants and the local contribution of a few pulsars with the latter expected to contribute more and more significantly as the energy increases.

\begin{figure}[ht]
\hskip -0.6 cm
\includegraphics[width=7.9cm,height=5.5cm]{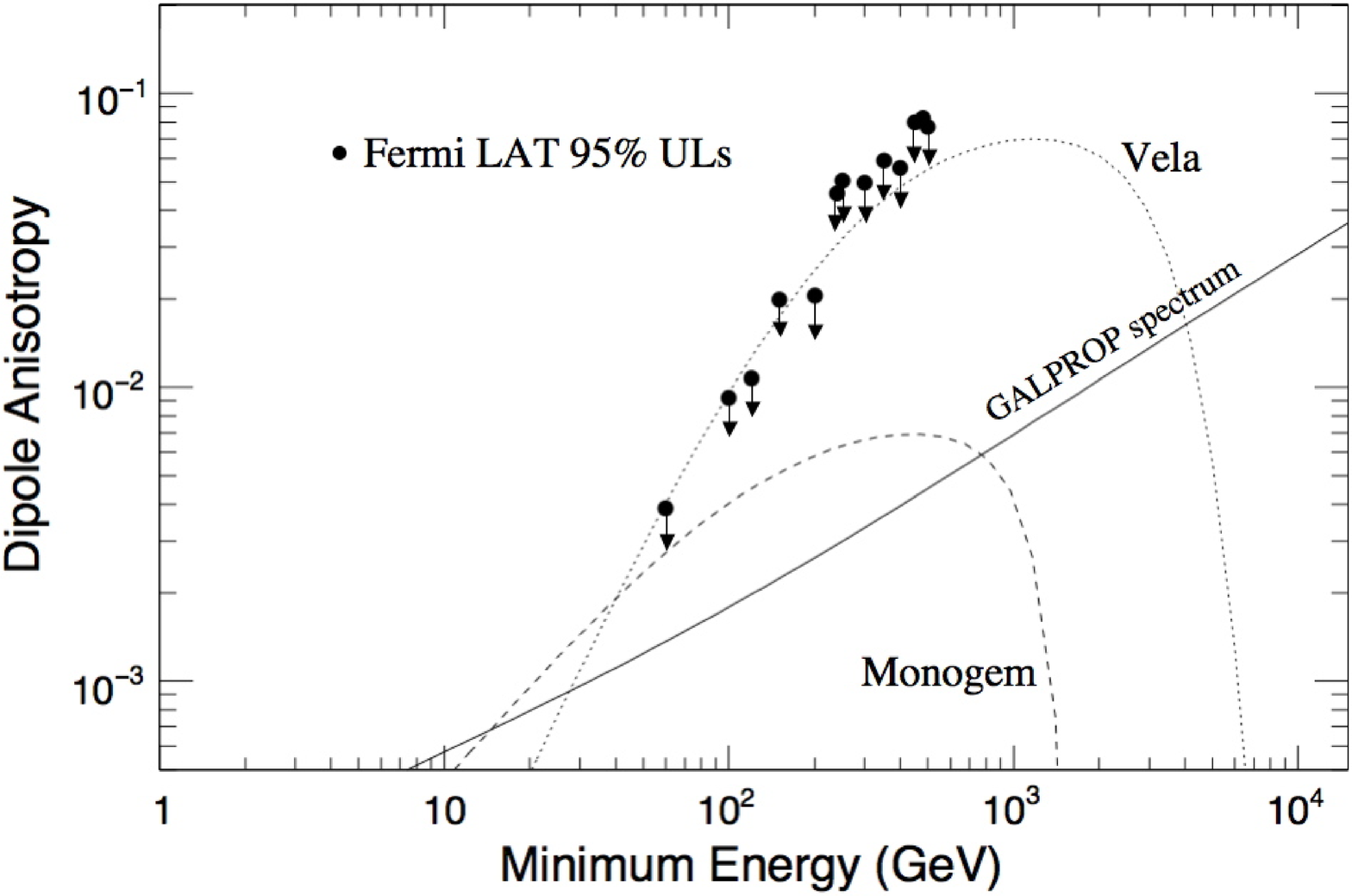}
\vskip -0.6 cm
\caption{\label{anisotropy_f} 
Dipole anisotropy $\delta$ versus the minimum energy for GALPROP (solid line), Monogem source (dashed line), and Vela source (dotted line). The 95\% Upper limit's confidence level from the data is also shown with circles. The solar modulation was treated using the force-field approximation
with  modulation potential  $\Phi$=550~MV.
}  
\end{figure}

\begin{figure}[ht]
\includegraphics[width=7.9cm,height=7.2cm]{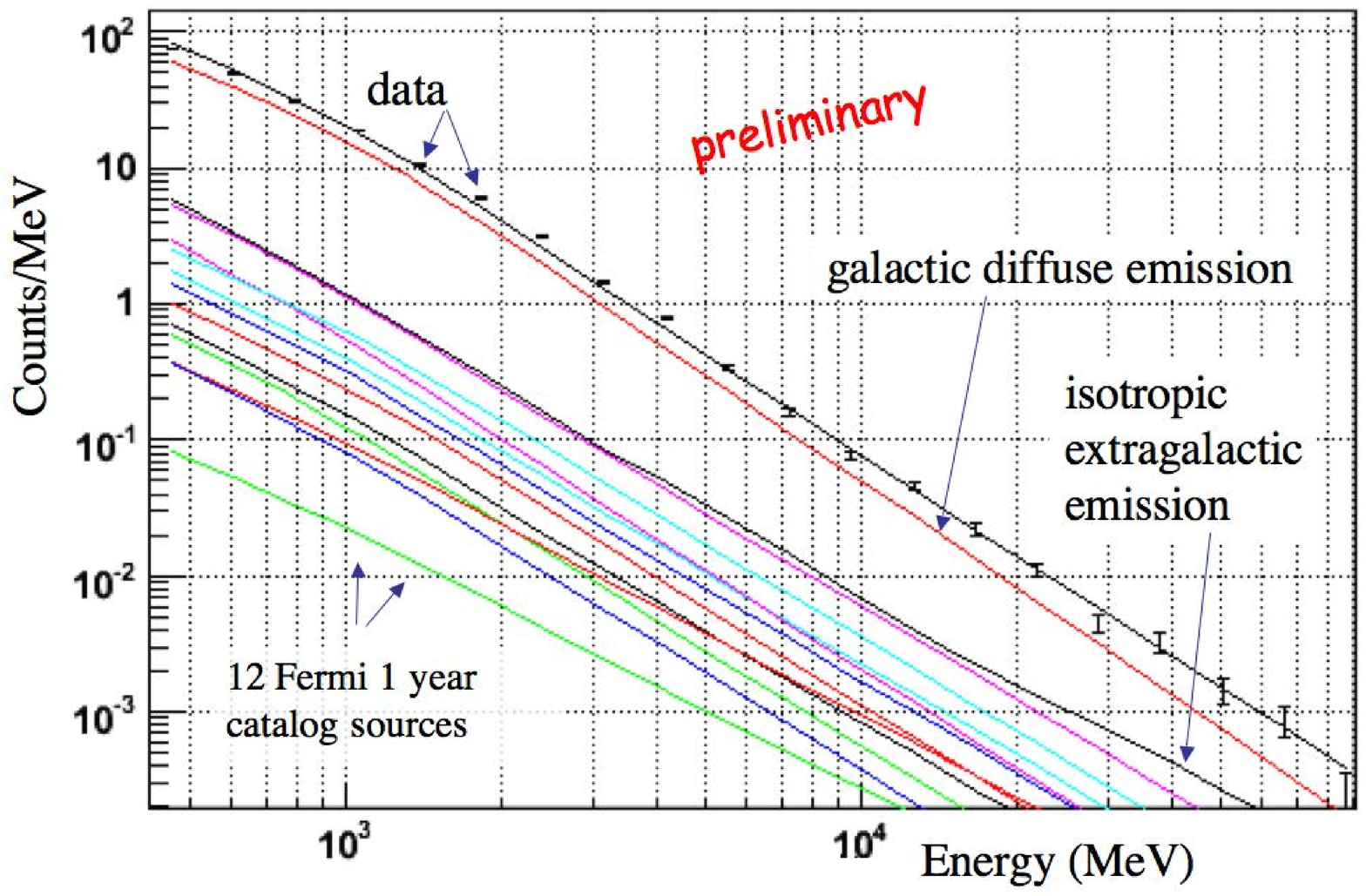}
\vskip -0.5 cm
\caption{ Counts spectra from the likelihood analysis of the $Fermi$ LAT data (number of counts vs reconstructed energy) in a 7$^{\circ} \times $7$^{\circ}$ region around the Galactic Center (number of counts vs reconstructed energy).
 }
\label{GC_fig}
\vskip -0.8 cm
\end{figure}
\begin{figure}[ht]
\includegraphics[width=7.9cm,height=5.5cm]{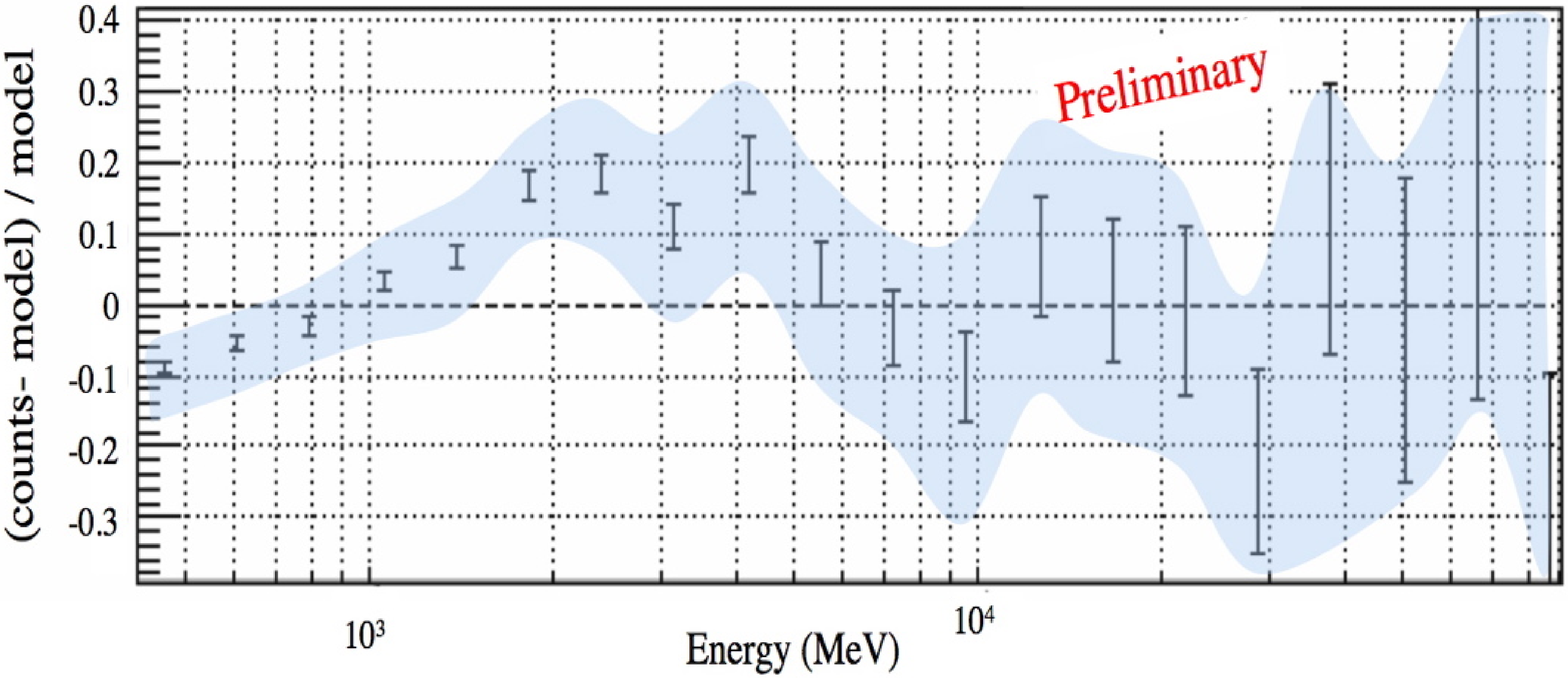}
\vskip -0.5 cm
\caption{ Residuals ( (exp.data - model)/model) of the above likelihood analysis. The blue area  shows the systematic errors on the effective area.}
\label{GC_r_fig}
\end{figure}
Two pulsars, Monogem,  at a distance of 290~pc and Geminga, at a distance of 160~pc,  can give a significant contribution to the high energy electron and positron flux reaching the Earth and with a set of reasonable parameters of the model of electron production the $Fermi$ LAT data and the PAMELA positron fraction can be well fit fraction~\cite{Pam_pos}  
(see figure  \ref{Fermi_el_fig}). However  we have a lot of freedom in the choice of these parameters because we still do not know much about these processes, so further study on high energy emission from pulsars is needed in order to confirm or reject the pulsar hypothesis.

Nevertheless a dark matter interpretation of the $Fermi$ LAT and of the PAMELA data is still an open possibility.  
 Figure \ref{fig:mod1}  shows the parameter space of dark matter particle mass versus pair-annihilation rate, for models where dark matter annihilates into monochromatic $e^\pm$  \cite{int_pap}.
The preferred range for the dark matter mass lies between 400 GeV and 1-2 TeV, with larger masses increasingly constrained by the H.E.S.S. results \cite{HESS}. The required annihilation rates, when employing a particular dark matter density profile  imply typical boost factors ranging between 20 and 100, when compared to the value $\langle\sigma v\rangle\sim3\times 10^{-26}\ {\rm cm}^3/{\rm sec}$ expected for a thermally produced dark matter particle relic.

How can one distinguish between the contributions of pulsars and dark matter annihilations? 
Most likely, a confirmation of the dark matter signal will require a consistency
between different experiments and new measurements of the reported excesses with large statistics.
The observed excess in the positron fraction should be consistent
with corresponding signals in absolute positron and electron fluxes in the PAMELA 
data and all lepton data collected by $Fermi$ LAT. 
$Fermi$ LAT has a large effective area and long projected lifetime, 5 years nominal with a 10 years goal, which makes it an excellent detector of cosmic-ray electrons up to $\sim$1 TeV. 
Future $Fermi$ LAT measurements of the total lepton flux with large statistics will enable distinguishing a gradual change in slope as opposed to a sharp cutoff with high confidence \cite{dark2}. The latter
 can be an indication in favor of the dark matter hypothesis. 

Another possibility is to look for anisotropies in the arrival directions of  the electrons. 
The  $Fermi$ LAT) detected more 
than 1.6 million 
cosmic-ray electrons/positrons with energies above 60~GeV during its first year of operation. 
The arrival directions of these events were searched for anisotropies 
of angular scale extending from $\sim$10$^\circ$ up to 90$^\circ$, and of 
minimum energy extending from 60~GeV up to 480~GeV. An upper limit for the dipole anisotropy has been set to 0.5 - 10\% depending on the energy  \cite{anisotropy_p}.

The levels of anisotropy expected for Vela-like and Monogem-like sources (i.e. sources with similar distances and ages) seem to be greater than the scale of anisotropies excluded by the results (see figure \ref{anisotropy_f} ). However, it is worth to point out that the model results are affected by large uncertainties related to the choice of the free parameters.

\section{ The gamma-ray signals}

A strong leptonic signal should be accompanied
by a boost in the $\gamma$-ray yield providing a distinct spectral signature
detectable by $Fermi$ LAT.

The Galactic center (GC) is expected to be the strongest source of $\gamma$-rays from DM 
annihilation, due to its coincidence with the cusped part of the DM halo density profile \cite{dark1},  \cite{dark}.
A preliminary analysis of the data, taken during the first 11 months of the Fermi satellite operations, is shown in figures  \ref{GC_fig} and \ref{GC_r_fig}.
The reported results were obtained with a binned  likelihood analysis, performed by means of the tools developed by the $Fermi$ LAT collaboration (gtlike, from the Fermi analysis tools \cite{tools}).  

In order to analyze the diffuse and point-source gamma-ray emission in this region using a maximum likelihood method for the LAT data, a  model of the already known sources and the diffuse  background should be built.
The model in use for the presented analysis contains   11 sources in the $Fermi$  1 year catalog \cite{Fermi_catalog} which are located  within or very close to  the considered region being analyzed.
These sources have a point-like spatial model and a spectrum in the form of a power-law.
The model also contains the diffuse gamma-ray background which is made of two components:
1)  {\sl the Galactic Diffuse gamma-ray background}. The observed  Galactic Diffuse emission was modeled  by means of the GALPROP code (model number 87XexphS)  \cite{str} and \cite{str2}.
\begin{figure}[ht]
\hskip -0.4 cm
\includegraphics[width=20pc, height=14pc]{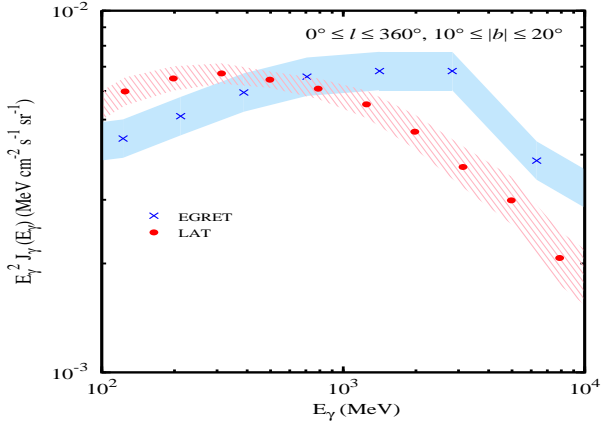}
\vskip -0.7 cm
\caption{\label{diffuse} 
 Diffuse emission intensity averaged over all Galactic 
longitudes for latitude range $10^\circ \leq |b| \leq 20^\circ$. Data points: 
$Fermi$ LAT, red dots; EGRET, blue crosses. Systematic uncertainties: $Fermi$ LAT, red; EGRET, blue. }
\end{figure}
\begin{figure}[ht]
\hskip -0.4 cm
\includegraphics[width=19pc, height=14pc]{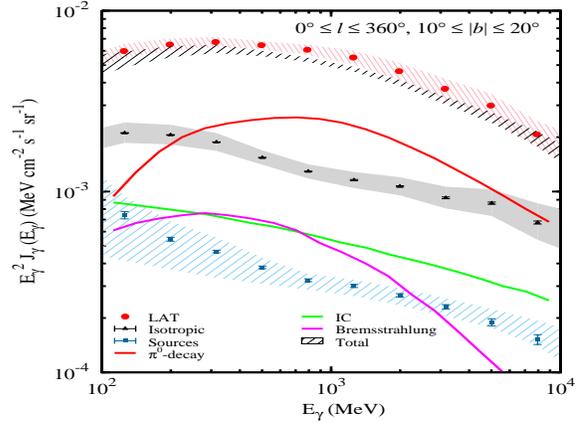}
\vskip -0.7 cm
\caption{\label{diffuse2} 
$Fermi$ LAT data with model, source, and isotropic components for  same sky region of Figure {\ref{diffuse}}}
\vskip -0.5 cm
\end{figure}
2)  {\sl the Isotropic Background}. This component should account for both the Extragalactic gamma-ray emission and  residual charged particles. It is modeled as an isotropic emission with a
template spectrum.

The diffuse gamma-ray backgrounds and discrete sources, as we know them today, can account for the large majority of the 
detected gamma-ray emission from the Galactic Center. Nevertheless a residual  emission is left, not accounted for by the above models \cite{F_sym}.

Improved modeling of the Galactic 
diffuse model as well as the potential contribution from other 
astrophysical sources (for instance unresolved point sources) could 
provide a better description of the data. Analyses are underway to 
investigate these possibilities.

An excess in gamma-ray  from dark matter annihilation also should be seen in the Galactic diffuse spectrum.
Figure~\ref{diffuse} shows  
the LAT data averaged over 
 all Galactic longitudes
and the latitude range 
$10^\circ \leq |b| \leq 20^\circ$.
The hatched band surrounding the LAT data indicates the systematic 
uncertainty in the measurement due to the uncertainty in the effective area.
Also shown on the right are  
the EGRET data for the same region 
of sky   where one can see that
the LAT-measured spectrum is significantly softer than the EGRET measurement \cite{FermiDiffuse}.
Figure~\ref{diffuse2} compares the LAT spectrum with the 
spectra of an {\em a priori} diffuse Galactic emission (DGE) model.
While the LAT spectral shape is consistent with the DGE model used in this 
paper, the overall model emission is too low thus giving rise to a 
$\sim10-15$\% excess over the energy range 100 MeV to 10 GeV.
However, the DGE model is based on pre-$Fermi$ LAT data and knowledge
of the DGE.
The difference between the model and data is of the same order as the 
uncertainty in the measured CR nuclei spectra at the relevant 
energies. Overall, 
the agreement between the LAT-measured spectrum and the model shows that 
the fundamental processes are consistent with our data, thus providing a
solid basis for future work in understanding the DGE.

\begin{figure}[ht]
\includegraphics[width=7.9cm,height=6.1cm]{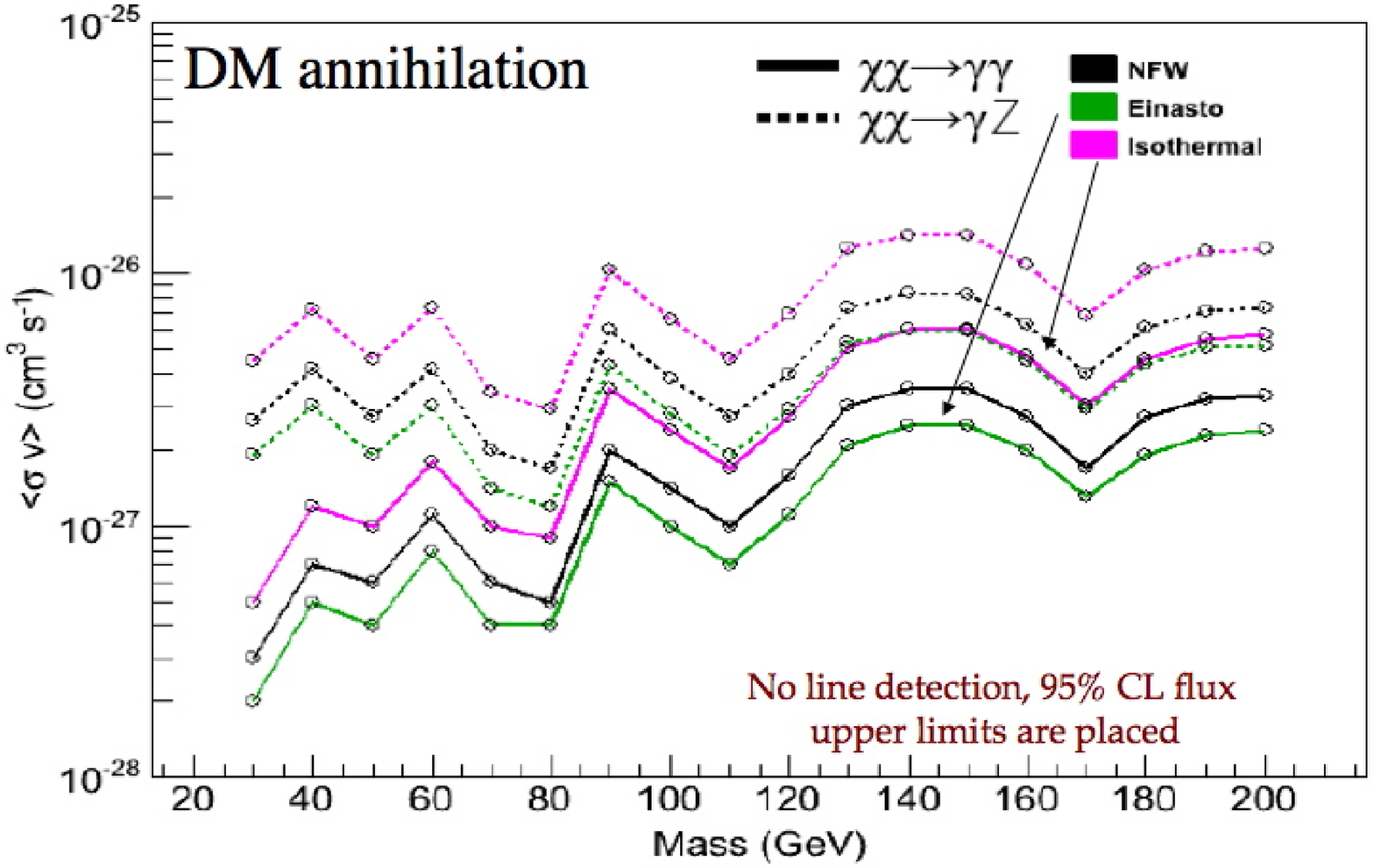}
\vskip -0.7 cm
\caption{\label{lines} 
Cross-section limits  for various  dark matter halo profiles for the  annihilation into monochromatic gamma-rays.}
\vskip -0.3 cm
\end{figure}
\begin{figure}[ht]
\includegraphics[width=7.9cm,height=6.1cm]{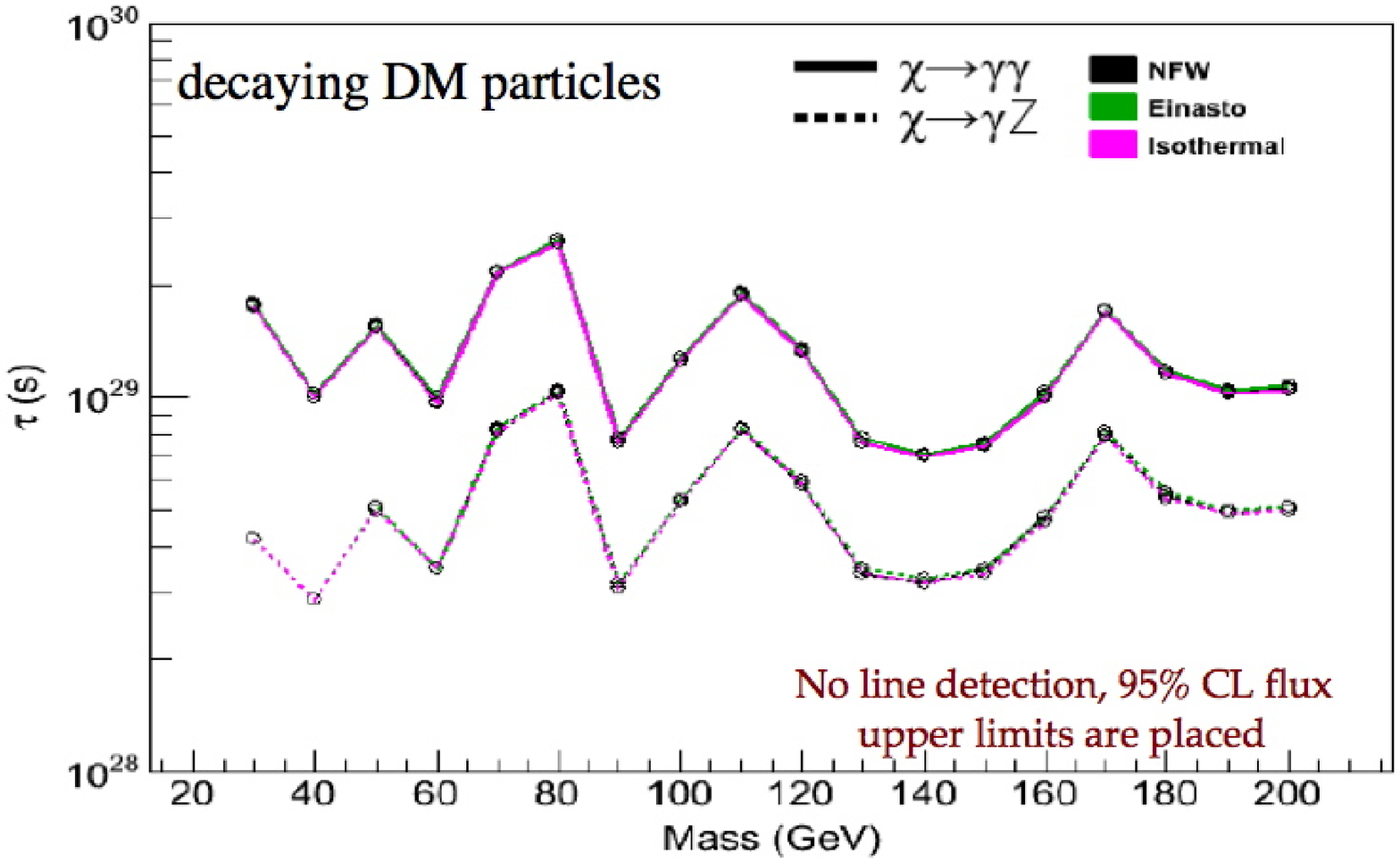}
\vskip -0.7 cm
\caption{\label{lines_decay} 
Lifetime  limits  for various  dark matter halo profiles for the decay channel into monochromatic gamma rays. }
\vskip -0.3 cm
\end{figure}

 \section{Dwarf spheroidal galaxies and Clusters  of galaxies}

  Local Group dwarf spheroidal galaxies, the largest galactic 
substructures predicted by the cold dark matter scenario, are attractive targets for dark matter indirect searches 
 because they are nearby and among the most extreme dark matter dominated environments.  
With the data taken during the first 11 months  no significant $\gamma$-ray emission was detected above 100 MeV from any dwarf galaxies. So we can determine upper limits to the $\gamma$-ray flux assuming both power-law spectra and representative spectra from WIMP annihilation. The resulting integral flux above 100 MeV is constrained to be at a level below around $10^{-9}$ photons cm$^{-2}$s$^{-1}$  \cite{Dwarf}.
Using recent stellar kinematic data, the $\gamma$-ray flux limits can be combined
  with improved determinations of the dark matter density profiles in 8 of the 14 candidate dwarfs to place limits on the pair annihilation
  cross-section of WIMPs in several widely studied
  extensions of the standard model, including its supersymmetric extension and other models that received recent attention.  
With the present data we are able to rule out
large parts of the parameter space where the thermal relic density is
below the observed cosmological dark matter density and WIMPs (neutralinos here) are
dominantly produced non-thermally, e.g. in models where supersymmetry
breaking occurs via anomaly mediation.
These $\gamma$-ray limits  also  
constrain some WIMP models proposed to explain the $Fermi$ LAT and PAMELA
  $e^+e^-$ data, including low-mass wino-like neutralinos and models with TeV masses pair-annihilating into muon-antimuon pairs.
The same kind of analysis can be made for  the clusters of galaxies \cite{Clusters}.

Finally a line at the WIMP mass, due to the 2$\gamma$ production channel, could be observed as a feature in the astrophysical source spectrum \cite{dark2}. Such an  observation would be  a ``smoking gun'' for WIMP DM as it is difficult to explain by a process other than WIMP annihilation or decay and the presence of a feature due to annihilation into 
$\gamma Z$ in addition would be even more convincing. 

Up to now however no lines vave been observed and we obtain $\gamma$-ray line flux upper limits in the range $0.6-4.5\times 10^{-9}\mathrm{cm}^{-2}\mathrm{s}^{-1}$  \cite{lines} and corresponding DM annihilation cross-section and decay lifetime limits shown in  figures \ref{lines}  and  \ref{lines_decay} .

\section{Conclusion}

Fermi Gamma-ray Space Telescope has opened a new era in DM searches and a large variety of analyses have been developed for clusters of galaxies, DM satellites, DM subhalos, cosmological DM and spectral lines. No significant detections have been made, but constraints that start to probe the available phase space have been put on the annihilation cross-section and decay lifetimes. In addition, several ongoing analyses are now being finalized, including studies of   the complicated galactic center region.

The CRE spectrum measured by $Fermi$ LAT is significantly harder than what was expected on the basis of previous data.  Adopting the presence of an extra $e^\pm$ primary component with $\sim$ 2.4 spectral index and $E_{cut} \sim 1$ TeV   allows a consistent interpretation of the $Fermi$ LAT CRE data, HESS and PAMELA.
Such an extra-component can be produced by nearby pulsars for a reasonable choice of relevant parameters
or by annihilating dark matter for models with $M_{DM}  \sim$ 1 TeV.
Improved analysis and complementary observations 
(CRE anisotropy, spectrum and angular distribution of diffuse $\gamma$,  DM sources search in $\gamma$) are required to possibly discriminate the right scenario.  
The dark matter origin of any exotic signal  has to be confirmed by complementary findings in $\gamma$-rays by $Fermi$ LAT and
atmospheric Cherenkov telescopes, and by LHC in the debris of high-energy proton destructions. 
On the other hand,  if the signal is due to to be a conventional astrophysical source of cosmic rays,
it will mean a direct detection of particles
accelerated at an astrophysical source, again a major breakthrough. 
However,
independent of the origin
of these excesses, exotic or conventional, we can expect a very exciting several years ahead of us.

\section{Acknowledgments}
The $Fermi$ LAT Collaboration acknowledges support from a number of agencies and institutes for both development and the operation of the LAT as well as scientific data analysis. These include NASA and DOE in the United States, CEA/Irfu and IN2P3/CNRS in France, ASI and INFN in Italy, MEXT, KEK, and JAXA in Japan, and the K.~A.~Wallenberg Foundation, the Swedish Research Council and the National Space Board in Sweden. Additional support from INAF in Italy and CNES in France for science analysis during the operations phase is also gratefully acknowledged.

\end{document}